\documentclass[11pt]{article}
\usepackage{amssymb}
\usepackage{amsfonts}
\usepackage{amsmath}
\usepackage{epsfig}
\usepackage{graphicx}
\usepackage{caption}
\usepackage{subcaption}

\begin{document}

\title{Brownian and fractional polymers with self-repulsion}
\author{Samuel Eleut\'{e}rio\thanks{%
CFTP, Instituto Superior T\'{e}cnico, Universidade de Lisboa,
sme@tecnico.ulisboa.pt} and R. Vilela Mendes\thanks{%
CMAFcIO, Faculdade de Ci\^{e}ncias, Universidade de Lisboa,
rvilela.mendes@gmail.com, rvmendes@fc.ul.pt,
https://label2.tecnico.ulisboa.pt/vilela/}}
\date{ }
\maketitle

\begin{abstract}
Brownian and fractional processes are useful computational tools for the
modelling of physical phenomena. Here, modelling linear homopolymers in
solution as Brownian or fractional processes, we develop a formalism to take
into account both the interactions of the polymer with the solvent as well
as the effect of arbitrary polymer-polymer potentials and Gibbs factors. As
an example the average squared length is computed for a non-trivial Gaussian
Gibbs factor, which is also compared with the Edwards' and a step factor.
\end{abstract}

\section{Brownian and fractional processes as a modelling and computational
tool}

From its discovery in the nineteen century \cite{Brown} to a model for
financial market fluctuations \cite{Bachelier}, to a clear evidence for the
existence of atoms \cite{Einstein1} and the inclusion in a more general
class of L\'{e}vy processes, Brownian motion has played a central role in
the modelling of physical phenomena and, in mathematics, in the development
of the field of stochastic and infinite-dimensional analysis \cite%
{StreitBook}. Brownian motion has independent increments, but is a
particular case of fractional Gaussian processes which, being either
positively or negatively correlated, allow for the modelling of memory
effects .

Brownian motion, and more general Gaussian processes, are also an useful
tool even in non-stochastic contexts. New solutions for deterministic
nonlinear partial differential equations may be obtained by a stochastic
construction, based on diffusion plus branching processes. This has been
used, for example, for the Navier-Stokes equation \cite{Jan} \cite{Waymire},
for magnetohydrodynamics \cite{FlorianiJCP}, for fractional \cite{Ouerdiane}
and singular partial differential equations \cite{StochSing}, etc.

Here we address the problem of modelling of polymers in solvents. The
behavior of polymers in solvents is determined by their chemical structures.
Very diverse patterns are observed. If, in a few instances, polymer
molecules display little interaction with the solvent, in most cases they
either curl up on itself or stretch out. A widely used characterization of
the polymer behavior is the Flory-Huggins parameter $\chi $ \cite{Flory1} 
\cite{Huggins1} which, based on the entropy of mixing, attempts to take into
account both the interaction of the polymer chains with the solvent
molecules and the polymer-polymer interaction. When the polymer is modeled
by a Gaussian process, the persistent nature of the fractional processes for 
$H>0.5$ and anti-persistent for $H<0.5$, allows the Hurst coefficient $H$ to
code for the polymer-solvent interaction, leaving the polymer-polymer
interaction to be modelled by diverse interaction potentials or Gibbs
factors.

The plan of the paper is the following: In Section 2 and 3, the polymer is
modelled as a Gaussian (Brownian, $H=0.5$ or fractional, $H\neq 0.5$)
process, each monomer being identified as a unit of "time" in the process.
Polymer-polymer interactions are taken into account by a potential $\Phi $
and its associated Gibbs factor. A general formula is obtained (Eqs.\ref{2.7}
and \ref{2.8}) to compute the average squared length for arbitrary
potentials through weighed integrals over Gaussian measures. Other physical
polymer variables may be obtained in a similar manner.

In Section 4 we discuss several types of potentials (and Gibbs factors), in
particular a potential that leads to a non-trivial Gaussian Gibbs factor and
a step factor. They are compared with the Edwards' factor \cite{Edwards},
which turns out to be identical to the step factor. For one of the Gibbs
factors, the one corresponding to the potential in Eq.(\ref{3.4}), an exact
analytical result is obtained for the average squared length (Eq.\ref{GF2}).
The average squared length $\mathbb{E}\left[ x^{2}\left( L\right) \right] $
is then computed over an extensive range of Hurst parameter values $H$ and
potential ranges (Fig.\ref{GaussFlory}). The main conclusions are:

- $\mathbb{E}\left[ x^{2}\left( L\right) \right] $ converges rapidly to a
power law in $L$, Fig.\ref{GaussR2}.

- The scaling index $\nu $ in $\sqrt{\mathbb{E}\left[ x^{2}\left( L\right) %
\right] }\sim L^{\nu }$, growths with $H$ and has an appreciable dependence
on the potential range.

- The Flory \cite{Flory} value for $\nu $ at $H=0.5$, obtained by scaling
arguments, roughly corresponds to a potential range of one monomer.

\section{Polymers in solvents as Gaussian processes}

Polymers are long chains of molecules (monomers) connected to each other at
points which allow for spatial rotations. Insofar as these rotations are
relatively unconstrained, polymers may be modelled by random walks or, in
the continuum limit, by paths of stochastic processes. However such models
should take into account the fact that, if two chain elements come close
together, molecular forces prevent them from occupying the same place.
Self-crossings of paths should be suppressed and this "excluded volume"
effect explicitly taken into account. On the other hand, swelling or
shrinking of the polymer chains depends on the nature of the solvent. This
suggests, for stochastic polymer models, the inclusion of both excluded
volume interactions and fractionality on the paths to model long range
correlations.

In the continuum version of the models, both in the Brownian and the
fractional Brownian cases, expectations are taken in the white noise measure 
\cite{Biagini} \cite{Mishura}, defined by%
\begin{equation}
\int_{S^{\prime }\left( R^{n}\right) }\exp i\left( \omega ,f\right) d\mu
\left( \omega \right) =\exp \left( -\frac{1}{2}\left\Vert f\right\Vert
_{L^{2}\left( R^{n}\right) }^{2}\right) \hspace{1cm}f\in S\left(
R^{n}\right) ,  \label{1.1}
\end{equation}%
Brownian motion being%
\begin{equation}
B\left( t\right) =\left\langle \omega |\chi _{\left[ 0,t\right]
}\right\rangle  \label{1.2a}
\end{equation}%
with%
\begin{equation}
\chi _{\left[ 0,t\right] }\left( s\right) =\left\{ 
\begin{array}{ccc}
1 &  & \text{if }0\leq s\leq t \\ 
0 &  & \text{otherwise}%
\end{array}%
\right.  \label{1.2b}
\end{equation}%
and fractional Brownian motion%
\begin{equation}
B^{H}\left( t\right) =\left\langle \omega |M_{\left[ 0,t\right]
}^{(H)}\right\rangle =\int_{0}^{t}M_{\left[ 0,t\right] }^{(H)}\left(
s\right) d\omega \left( s\right)  \label{1.3a}
\end{equation}%
where here, for computational convenience, we use the Molchan representation
via the Wiener process on a finite time interval \cite{Norros} \cite%
{Mishura2}%
\begin{equation}
M_{t}^{(H)}\left( s\right) \equiv M_{\left[ 0,t\right] }^{(H)}\left(
s\right) =c_{H}\left\{ \left( \frac{t}{s}\right) ^{H-\frac{1}{2}}\left(
t-s\right) ^{H-\frac{1}{2}}-\left( H-\frac{1}{2}\right) s^{\frac{1}{2}%
-H}\int_{s}^{t}u^{H-\frac{3}{2}}\left( u-s\right) ^{H-\frac{1}{2}}du\right\}
\label{1.3b}
\end{equation}%
and%
\begin{equation}
c_{H}=\left( \frac{2H\Gamma \left( \frac{3}{2}-H\right) }{\Gamma \left( H+%
\frac{1}{2}\right) \Gamma \left( 2-2H\right) }\right) ^{\frac{1}{2}}
\label{1.3c}
\end{equation}

In the continuum model, to mimic the excluded volume effect, Gibbs factors
are included in the path integral. In the Edwards model, for example, the
factor is%
\begin{equation}
G=\frac{1}{Z}\exp \left( -g\int_{0}^{L}dt\int_{0}^{t}ds\delta \left(
\left\vert B^{H}\left( t\right) -B^{H}\left( s\right) \right\vert \right)
\right)  \label{1.3}
\end{equation}%
with $H=\frac{1}{2}$ in the original Edwards model \cite{Edwards} and $H\neq 
\frac{1}{2}$ in the fractional generalizations \cite{Grothaus}. The Gibbs
factor $G$ means that the excluded volume effect is controlled by the
self-intersection local time $\mathcal{L}$ of the process,%
\begin{equation}
\mathcal{L}=\int_{0}^{L}dt\int_{0}^{t}ds\delta \left( \left\vert B^{H}\left(
t\right) -B^{H}\left( s\right) \right\vert \right) .  \label{1.4}
\end{equation}

Conceptually, this is an elegant model. However, due to the nature of $%
\mathcal{L}$, the physical interpretation of the Gibbs factor in (\ref{1.3})
is rather delicate. When the $\delta $ distribution in (\ref{1.4}) is
approximated by a sequence of functions, for example%
\begin{eqnarray}
\delta _{\varepsilon }\left( x\right) &=&\frac{1}{\left( 2\pi \varepsilon
\right) ^{n/2}}e^{-\frac{x^{2}}{2\varepsilon }};\;x\in \mathbb{R}%
^{n};\;\varepsilon >0,  \notag \\
\mathcal{L}_{\varepsilon } &=&\int_{0}^{L}dt\int_{0}^{t}ds\delta
_{\varepsilon }\left( \left\vert B^{H}\left( t\right) -B^{H}\left( s\right)
\right\vert \right) ,  \label{1.5}
\end{eqnarray}%
one finds out that for $H=\frac{1}{2}$, $\mathcal{L}_{\varepsilon }$
converges in $L^{2}$ only for $n=1$, $n$ being the spatial dimension of the
space. For $n=2$ an infinite quantity must be subtracted, centering the
distribution \cite{Vara}%
\begin{equation*}
\mathcal{L}_{\varepsilon ,c}=\mathcal{L}_{\varepsilon }-\mathbb{E}\left( 
\mathcal{L}_{\varepsilon }\right) .
\end{equation*}%
In $n=2$ this is sufficient to ensure convergence of (\ref{1.5}) when $%
\varepsilon \searrow 0$. For $n\geq 3$ a further multiplicative factor is
needed to yield a (in law) limiting process \cite{Yor} \cite{Calais}. For
the fractional case similar renormalization procedures have been found \cite%
{Grothaus} \cite{Hu1} \cite{Hu2} \cite{Oliveira}.

All this gave rise to some pieces of beautiful mathematics but, as far as
polymer physics is concerned, the model only considers the polymers as paths
without transversal dimensions (which they have) and the molecular
repellency as a pointwise intersection without any particular information on
the molecular potential. In addition, the need for renormalization
procedures or subtraction of divergent quantities, removes any relation of
the results to the original coupling constants of the model.

Pointwise self interaction neglects effects of polymer structure on the
physical properties and complex functions \cite{Alberts} of the polymers.
This suggests that some other Gibbs factors, that also punish
self-intersection, might be used to obtain results which take in account the
polymer physical dimensions and, at the same time, rigorously relate the
results to the parameters in the model. In this paper an attempt is made in
this direction by developing a general analytical framework for the
exploration of arbitrary interaction potentials. As an example, the average
squared length is computed for a non-trivial Gaussian Gibbs factor, which is
also compared with the Edwards and step Gibbs factors.

\section{The average squared length}

Here, as in the Edwards model, the underlying probability space is
continuous white noise. However, for the calculation of expectation values,
a discrete time sampling of the path will be used. For polymers, a discrete
time sampling formulation is appropriate, each time step corresponding to a
monomer. Then, all expectations in a linear $L-$monomers homopolymer chain
will be obtained from measures%
\begin{equation}
d\nu \left( \omega \right) =\frac{1}{Z}d\mu \left( \omega \right)
\prod_{\alpha ^{\prime }>\alpha \geq 1}^{L}e^{-\Phi \left( \left\vert
\left\langle \omega |M_{\alpha }^{(H)}\right\rangle -\left\langle \omega
|M_{\alpha ^{\prime }}^{(H)}\right\rangle \right\vert \right) }  \label{2.1}
\end{equation}%
where $\Phi \left( \left\vert \left\langle \omega |M_{\alpha
}^{(H)}\right\rangle -\left\langle \omega |M_{\alpha ^{\prime
}}^{(H)}\right\rangle \right\vert \right) $ is the excluded volume potential
associated to interaction of the pair $\alpha ,\alpha ^{\prime }$ of polymer
segments. For the Brownian ($H=\frac{1}{2}$) case $M_{\alpha }^{(H)}$ is the
characteristic function (\ref{1.2b}) of the interval $\left[ 0,\alpha \right]
$ and for the fractional case the integration kernel (\ref{1.3b}-\ref{1.3c})
discussed in the introduction. $\omega =\left\{ \omega _{\alpha }\right\} $
is $n-$dimensional white noise and $\mu \left( \omega \right) $ the white
noise Gaussian measure. The Gibbs factors should take into account all the $%
M=\frac{L(L-1)}{2}$ ordered $\left( \alpha ,\alpha ^{\prime }\right) $ pairs
interactions. The partition function $Z$ is%
\begin{equation}
Z=\int_{S^{\prime n}}d\mu \left( \omega \right) \prod_{\alpha ^{\prime
}>\alpha \geq 1}^{L}e^{-\Phi \left( \left\vert \left\langle \omega
|M_{\alpha }^{(H)}\right\rangle -\left\langle \omega |M_{\alpha ^{\prime
}}^{(H)}\right\rangle \right\vert \right) }.  \label{2.2}
\end{equation}%
and the average squared length%
\begin{eqnarray}
\mathbb{E}\left[ x^{2}\left( L\right) \right]  &=&\frac{1}{Z}\int_{S^{\prime
n}}d\mu \left( \omega \right) \left\vert \left\langle \omega
|M_{L}^{(H)}\right\rangle \right\vert ^{2}\prod_{\alpha ^{\prime }>\alpha
\geq 1}^{L}e^{-\Phi \left( \left\vert \left\langle \omega |M_{\alpha
}^{(H)}\right\rangle -\left\langle \omega |M_{\alpha ^{\prime
}}^{(H)}\right\rangle \right\vert \right) }  \notag \\
&&  \label{2.3}
\end{eqnarray}%
with lengths defined in the $\mathbb{R}^{n}$ Euclidean metric and
integration of the Wiener process with the kernels $M_{\alpha }^{(H)}$. At
each evolution step in (\ref{2.3}) replace $\omega $ by the operator $\frac{1%
}{i}\frac{\partial }{\partial \overrightarrow{J}}$ acting on $\exp i\left(
\omega ,\overrightarrow{J}\right) |_{\overrightarrow{J}=0}$. $%
\overrightarrow{J}$ is a $nL-$dimensional vector function with components in 
$S\left( R^{n}\right) $. Then, integrating over the $S^{\prime }\left(
R^{n}\right) $ white noise measure for each one of the $L$ evolution steps,
one obtains%
\begin{eqnarray}
\mathbb{E}\left[ x^{2}\left( L\right) \right]  &=&\frac{1}{Z}\left\vert
\left\langle \frac{1}{i}\frac{\partial }{\partial \overrightarrow{J}}%
|M_{L}^{(H)}\right\rangle \right\vert ^{2}  \notag \\
&&\left. \exp \left\{ -\Phi \left( \left\vert \left\langle \frac{1}{i}\frac{%
\partial }{\partial \overrightarrow{J}}|M_{\alpha }^{(H)}\right\rangle
-\left\langle \frac{1}{i}\frac{\partial }{\partial \overrightarrow{J}}%
|M_{\alpha ^{\prime }}^{(H)}\right\rangle \right\vert \right) \right\} \exp
\left\{ -\frac{1}{2}\left\Vert \overrightarrow{J}\right\Vert ^{2}\right\}
\right\vert _{\overrightarrow{J}=0}.  \notag
\end{eqnarray}%
Now one uses%
\begin{equation}
\left. f\left( \frac{1}{i}\frac{\partial }{\partial \overrightarrow{J}}%
\right) e^{-\frac{1}{2}\left\Vert \overrightarrow{J}\right\Vert
_{n}^{2}}\right\vert _{\overrightarrow{J}=0}=\left( 2\pi \right) ^{-\frac{n}{%
2}}\int d^{n}\overrightarrow{y}_{\alpha }f\left( \overrightarrow{y}_{\alpha
}\right) e^{-\frac{1}{2}\left\Vert \overrightarrow{y}_{\alpha }\right\Vert
^{2}}  \label{2.6}
\end{equation}%
to obtain%
\begin{equation}
\mathbb{E}\left[ x^{2}\left( L\right) \right] =\frac{1}{Z}\int \left(
\prod_{\alpha =1}^{L}d^{n}y_{\alpha }e^{-\frac{1}{2}\sum_{\alpha }\left\vert 
\overrightarrow{y}_{\alpha }\right\vert ^{2}}\right) \left\vert \left\langle 
\overrightarrow{y}_{L}|M_{L}^{(H)}\right\rangle \right\vert
^{2}\prod_{\alpha ^{\prime }>\alpha \geq 1}^{L}e^{-\Phi \left( \left\vert
\left\langle \overrightarrow{y}_{\alpha }|M_{\alpha }^{(H)}\right\rangle
-\left\langle \overrightarrow{y}_{\alpha ^{\prime }}|M_{\alpha ^{\prime
}}^{(H)}\right\rangle \right\vert \right) }  \label{2.7}
\end{equation}%
and%
\begin{equation}
Z=\int \left( \prod_{\alpha =1}^{L}d^{n}y_{\alpha }e^{-\frac{1}{2}%
\sum_{\alpha }\left\vert \overrightarrow{y}_{\alpha }\right\vert
^{2}}\right) \prod_{\alpha ^{\prime }>\alpha \geq 1}^{L}e^{-\Phi \left(
\left\vert \left\langle \overrightarrow{y}_{\alpha }|M_{\alpha
}^{(H)}\right\rangle -\left\langle \overrightarrow{y}_{\alpha ^{\prime
}}|M_{\alpha ^{\prime }}^{(H)}\right\rangle \right\vert \right) }
\label{2.8}
\end{equation}%
where $2\pi $ powers are discarded both in $\mathbb{E}\left[ x^{2}\left(
L\right) \right] $ and $Z$.

At this point and before using Eqs.(\ref{2.7}) and (\ref{2.8}) for
calculations with diverse Gibbs factors, it is useful to ponder the meaning
of the replacement of the process $\omega $ in (\ref{2.3}) and (\ref{2.2})
by an integration over a Gaussian probability density in (\ref{2.7}) and (%
\ref{2.8}) and the use of an integration variable $\overrightarrow{y}%
_{\alpha }$ for each evolution step.

When computing physical quantities, the Gibbs potential assigns a weight
that is a function of the variable $\left\vert \left\langle \omega
|M_{\alpha }^{(H)}\right\rangle -\left\langle \omega |M_{\alpha ^{\prime
}}^{(H)}\right\rangle \right\vert $, that is, the separation of the process
at the "times" $\alpha $ and $\alpha ^{\prime }$. For a fractional process
in dimension $n$, with Hurst coefficient $H$, the probability $\rho \left( 
\overrightarrow{z}\right) $ for a separation $\overrightarrow{z}$ after $%
\left\vert \alpha -\alpha ^{\prime }\right\vert $ \ evolution steps is%
\begin{equation}
\rho \left( \overrightarrow{z}\right) =\left( \frac{1}{2\pi n\left\vert
\alpha -\alpha ^{\prime }\right\vert ^{2H}}\right) ^{\frac{n}{2}}\exp \left(
-\frac{1}{2}\frac{\left\vert \overrightarrow{z}\right\vert }{n\left\vert
\alpha -\alpha ^{\prime }\right\vert ^{2H}}\right)  \label{2.9}
\end{equation}%
the covariance being $n\left\vert \alpha -\alpha ^{\prime }\right\vert ^{2H}$%
. Examination of (\ref{2.6}) - (\ref{2.8}) and a change of variables%
\begin{equation*}
\overrightarrow{y}=\frac{\overrightarrow{z}}{\sqrt{n}\left\vert \alpha
-\alpha ^{\prime }\right\vert ^{H}}
\end{equation*}%
shows that what is being done is an integration on the separation
probabilities $\rho \left( \overrightarrow{z}\right) $.

\section{Gibbs factors}

Having taken explicit account of the relevant stochastic processes, Eqs.(\ref%
{2.7}) and (\ref{2.8}) are appropriate tools to explore, analytically or
numerically, the effect of several types of polymer self-interactions
corresponding to any type of interaction potential. The Edwards Gibbs factor
used in the past to obtain, modulo renormalization procedures, some results
for Brownian and fractional polymers (see \cite{Grothaus},\cite{Oliveira}
and references therein) was%
\begin{equation}
e^{-\phi _{Edwards}\left( x\right) }=e^{-g\delta \left( \left\vert
x\right\vert \right) }  \label{3.1}
\end{equation}%
Actually, the excluding volume effect of the Edwards' factor is very similar
to the one of a step factor%
\begin{equation}
e^{-\phi _{step}\left( x\right) }=\theta \left( \left\vert x\right\vert
-\varepsilon \right) .  \label{3.1a}
\end{equation}

This is easily seen, for example, by approximating the delta distribution by
a sequence of rectangular functions of constant area and computing the
Fourier transform%
\begin{equation*}
\widehat{f}_{Edwards}\left( q\right) =\left( 2\pi \right) ^{\frac{n}{2}%
}\delta ^{n}\left( q\right) -\lim_{\varepsilon \rightarrow 0}\frac{1}{%
\left\vert q\right\vert ^{\frac{n-2}{2}}}\int_{0}^{\varepsilon }r^{\frac{n}{2%
}}J_{\frac{n-2}{2}}\left( \left\vert q\right\vert r\right) dr\left( 1-e^{-%
\frac{g}{\varepsilon }}\right) .
\end{equation*}%
For the step Gibbs factor the Fourier transform would be%
\begin{equation*}
\widehat{f}_{step}\left( q\right) =\left( 2\pi \right) ^{\frac{n}{2}}\delta
^{n}\left( q\right) -\lim_{\varepsilon \rightarrow 0}\frac{1}{\left\vert
q\right\vert ^{\frac{n-2}{2}}}\int_{0}^{\varepsilon }r^{\frac{n}{2}}J_{\frac{%
n-2}{2}}\left( \left\vert q\right\vert r\right) dr
\end{equation*}%
the conclusion being that for small $\varepsilon $ the Edwards' Gibbs factor
is identical to a step factor and the coupling constant $g$ becomes an
irrelevant parameter. This is an useful fact, because it is much simpler to
do practical calculations with the step factor than to handle the
self-intersection local time (\ref{1.4})\footnote{%
It is also a step factor that is used in most numerical simulations
performed in the past for Brownian and fractional polymers.}.

\begin{figure}[tbh]
\centering
\includegraphics[width=0.8\textwidth]{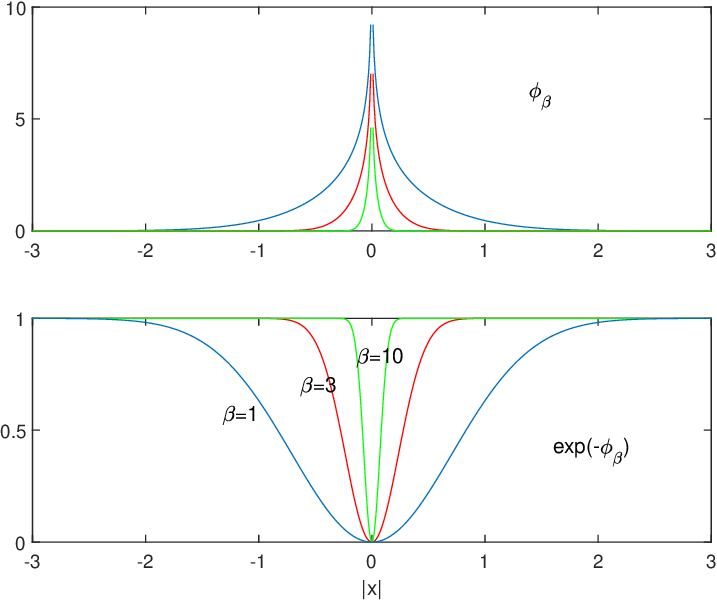}
\caption{Potential $\protect\phi $ and Gibbs factor $\exp \left( -\protect%
\phi \right) $ for several values of $\protect\beta $ ($\protect\beta =1,3,10
$)}
\label{Fig1}
\end{figure}

Another Gibbs factor for which explicit analytic results, without
renormalization procedures, are obtained, corresponds to the potential%
\begin{equation}
\phi _{\beta }\left( \left\vert x\right\vert \right) =\log \frac{1}{1-\exp
\left( -\beta ^{2}\left\vert x\right\vert ^{2}\right) }  \label{3.4}
\end{equation}%
with an associated Gibbs factor%
\begin{equation}
e^{-\phi _{\beta }\left( \left\vert x\right\vert \right) }=1-\exp \left(
-\beta ^{2}\left\vert x\right\vert ^{2}\right)   \label{3.2}
\end{equation}%
Fig.\ref{Fig1} displays the potential and the Gibbs factor for several
values of $\beta $ ($\beta =1,3,10$). Like the Edwards's factor, this one
also extracts from the integration domain all the $\left\vert x\right\vert =0
$ hyperplanes, that is, all hyperplanes corresponding to polymer
intersections but, depending on the parameter $\beta $, it also allows for
repulsion at non-zero distances.

The Fourier transform of this Gibbs factor is%
\begin{equation}
\widehat{f}\left( q\right) =\left( 2\pi \right) ^{\frac{n}{2}}\delta
^{n}\left( q\right) -\frac{1}{\left( 2\beta ^{2}\right) ^{\frac{n}{2}}}\exp
\left( -\frac{\left\vert q\right\vert ^{2}}{4\beta ^{2}}\right)  \label{3.3}
\end{equation}

For comparison with the step (or Edwards') factors one may compute the
amount $\gamma $ of integration volume of a sphere of radius $R$ that is
excluded by these factors,%
\begin{equation*}
\gamma =\frac{1}{V_{n}\left( R\right) }\int_{S\left( R\right) }e^{-\phi
}d^{n}x.
\end{equation*}%
For the step factor, identical to Edwards' for small $\varepsilon $, it is%
\begin{equation*}
\gamma \left( \varepsilon \right) =1-\left( \frac{\varepsilon }{R}\right)
^{n}
\end{equation*}%
and for the Gaussian factor (for large $R$)%
\begin{equation*}
\gamma \left( \beta \right) \simeq 1-\frac{\Gamma \left( \frac{n}{2}%
+1\right) }{\left( R\beta \right) ^{n}}.
\end{equation*}%
Therefore one expects qualitatively similar effects, at large $\beta $
(small $\varepsilon $), if $\beta \sim \frac{1}{\varepsilon }$.

\subsection{The Gaussian factor}

\begin{equation}
\mathbb{E}\left[ x^{2}\left( L\right) \right] _{G}=\frac{1}{Z}\int \left(
\prod_{\alpha =1}^{L}d^{n}y_{\alpha }e^{-\frac{1}{2}\sum_{\alpha }\left\vert 
\overrightarrow{y}_{\alpha }\right\vert ^{2}}\right) \left\vert \left\langle 
\overrightarrow{y}|M_{L}^{(H)}\right\rangle \right\vert ^{2}\prod_{\alpha
^{\prime }>\alpha \geq 1}^{L}\left( 1-e^{-\beta ^{2}\left\vert \left\langle
y|M_{\alpha }^{(H)}\right\rangle -\left\langle y|M_{\alpha ^{\prime
}}^{(H)}\right\rangle \right\vert ^{2}}\right)   \label{GF0}
\end{equation}%
\begin{equation}
Z=\int \left( \prod_{\alpha =1}^{L}d^{n}y_{\alpha }e^{-\frac{1}{2}%
\sum_{\alpha }\left\vert \overrightarrow{y}_{\alpha }\right\vert
^{2}}\right) \prod_{\alpha ^{\prime }>\alpha \geq 1}^{L}\left( 1-e^{-\beta
^{2}\left\vert \left\langle y|M_{\alpha }^{(H)}\right\rangle -\left\langle
y|M_{\alpha ^{\prime }}^{(H)}\right\rangle \right\vert ^{2}}\right) 
\label{GF0a}
\end{equation}

In the Gibbs factor the contribution corresponding to an interaction between
the evolution steps $\alpha $ and $\alpha ^{\prime }$ is%
\begin{eqnarray}
&&1-\exp \left( -\beta ^{2}\left\vert \left\langle y|M_{\alpha
}^{(H)}\right\rangle -\left\langle y|M_{\alpha ^{\prime
}}^{(H)}\right\rangle \right\vert ^{2}\right)  \label{GF1} \\
&=&1-\exp \left( -\beta ^{2}\left\vert \sum_{i=1}^{\alpha }\overrightarrow{y}%
\left( i\right) M_{\alpha }^{(H)}\left( i\right) -\sum_{i=1}^{\alpha
^{\prime }}\overrightarrow{y}\left( i\right) M_{\alpha ^{\prime
}}^{(H)}\left( i\right) \right\vert ^{2}\right)  \notag
\end{eqnarray}%
In the expressions above $y$ is a $nL-$dimensional vector and $M_{\alpha
}^{(H)}$ a $\alpha -$dimensional one (or a $L-$dimensional vector, where
only the first $\alpha $ elements are nonzero). $M_{\alpha }^{(H)}\left(
i\right) $ and $M_{\alpha ^{\prime }}^{(H)}\left( i\right) $ are discretized
versions of the integration kernel, computed at the middle point of each
interval%
\begin{equation*}
M_{\alpha }^{(H)}\left( i\right) =c_{H}\left\{ \left( \frac{\alpha }{i-\frac{%
1}{2}}\right) ^{H-\frac{1}{2}}\left( \alpha -i-\frac{1}{2}\right) ^{H-\frac{1%
}{2}}-\left( H-\frac{1}{2}\right) \left( i-\frac{1}{2}\right) ^{\frac{1}{2}%
-H}\int_{i-\frac{1}{2}}^{\alpha }u^{H-\frac{3}{2}}\left( u-i-\frac{1}{2}%
\right) ^{H-\frac{1}{2}}du\right\}
\end{equation*}

\begin{figure}[tbh]
\centering
\begin{subfigure}{.49\textwidth}
  \centering
  \includegraphics[width=1\textwidth]{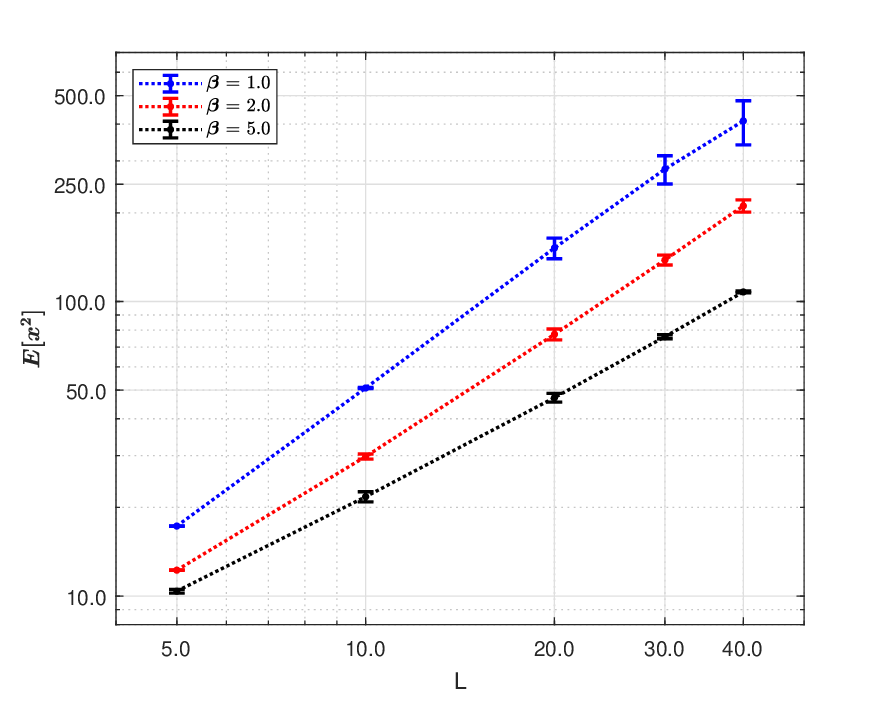}
  \caption{Dimension 2}
  \label{GaussR2Dim2}
\end{subfigure}
\begin{subfigure}{.49\textwidth}
  \centering
  \includegraphics[width=1\textwidth]{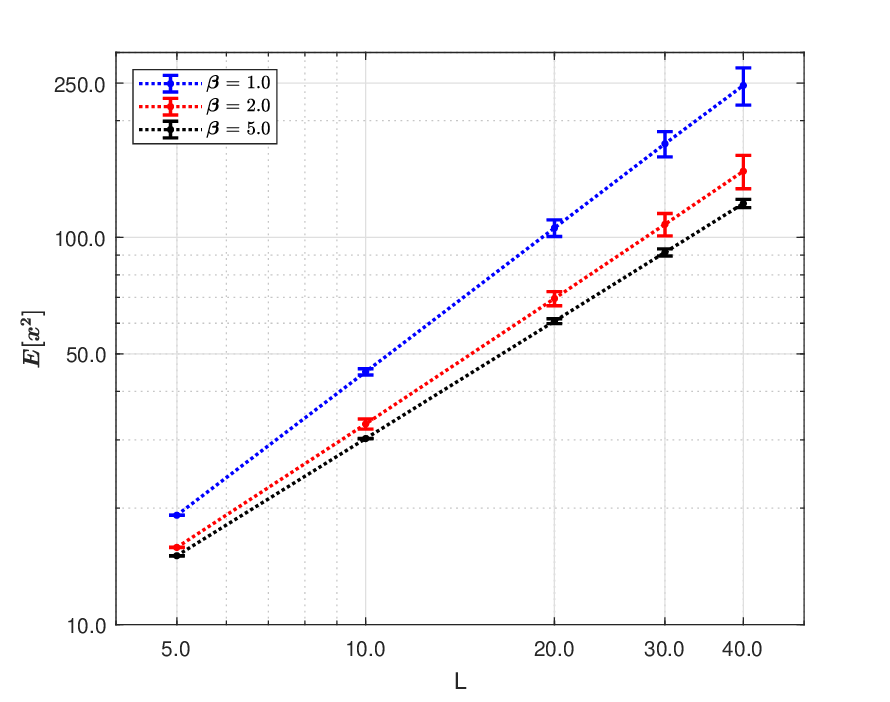}
  \caption{Dimension 3}
  \label{GaussR2Dim}
\end{subfigure}
\caption{The power law dependence of the average squared length for H=0.5}
\label{GaussR2}
\end{figure}

Because of the Gaussian measure the integrals in (\ref{GF0}) and (\ref{GF0a}%
) are easily computed by importance sampling displaying an accurate power
law behavior even for relatively small $L$ values (see Fig.\ref{GaussR2}).

This power law is quite evident in the figure, for three different values of
the constant $\beta $ ($1.0,2.0$,$5.0$ and $H=0.5$). Notice that small
values of $\beta $ correspond to situations of large phase space volume
exclusion, while large values correspond to little or no exclusion. When the
integrals in (\ref{GF0}) and (\ref{GF0a}) are computed, as the value of $%
\beta $ decreases an increase in the error is expected, due to the increase
in the weight of the exclusion. The same occurs when the length $L$
increases, due to the significant increase in the space over which the
integration takes place.

\begin{figure}[tbh]
\centering%
\begin{subfigure}{.49\textwidth}
  \centering
  \includegraphics[width=1\textwidth]{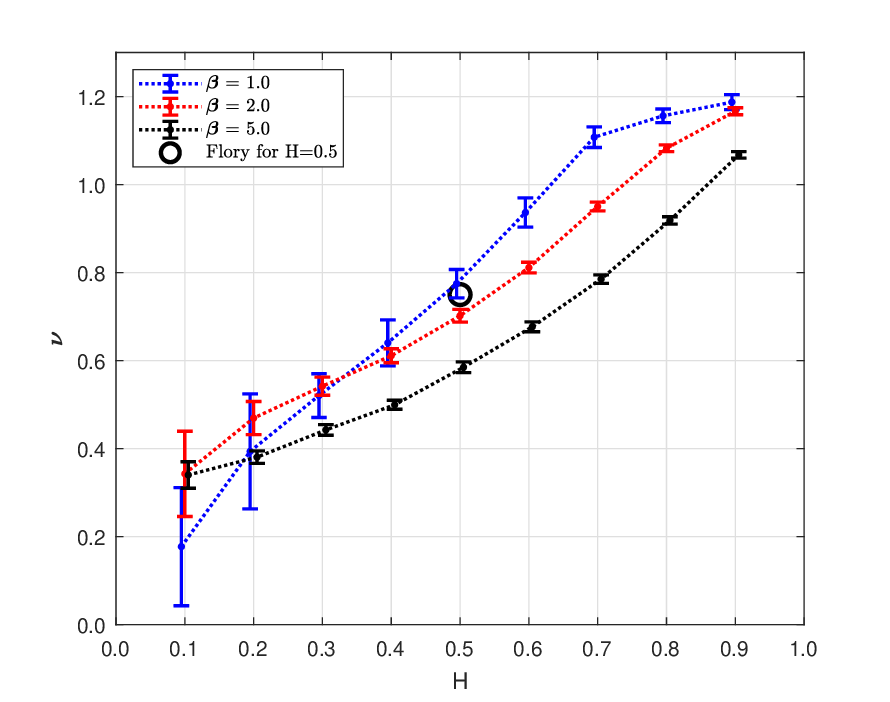}
  \caption{Dimension 2}
  \label{GaussFloryDim2}
\end{subfigure}
\begin{subfigure}{.49\textwidth}
  \centering
  \includegraphics[width=1\textwidth]{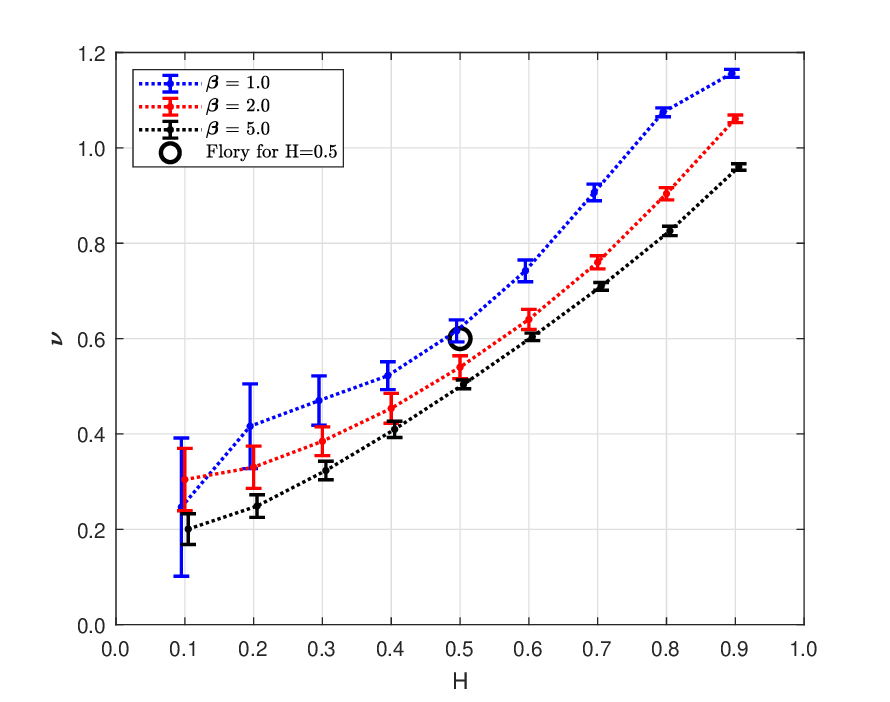}
  \caption{Dimension 3}
  \label{GaussFloryDim3}
\end{subfigure}
\caption{The scaling index}
\label{GaussFlory}
\end{figure}

Of most physical relevance is the asymptotic slope $\nu $ of%
\begin{equation}
\sqrt{\mathbb{E}\left[ x^{2}\left( L\right) \right] }\sim L^{\nu }
\label{GF8}
\end{equation}%
Fig.\ref{GaussFlory} shows results\footnote{%
Estimated from $L$ in the range $10-40.$} for dimensions $n=2$ and $n=3$.
Error bars in the figure account for statistical errors in the computation
of the integrals. The important facts to retain are both the growth of the $%
\nu $ values\ with $H$ and their dependency on the range of the potential.
The Flory \cite{Flory} points ($\nu =0.75$ for $n=2$ and $\nu =0.6$ at $%
H=0.5 $) are marked with \textquotedblright $\bigcirc $\textquotedblright\
on the plots. They correspond to the situation where the average range of
the potential is close to the size of a monomer (a discrete evolution step
in our stochastic representation).

Because all integrals in (\ref{GF0}-\ref{GF0a}) are Gaussian a closed form
expression for $\mathbb{E}\left[ x^{2}\left( L\right) \right] _{G}$ may in
this case be obtained. First one expands the $\prod_{\alpha ^{\prime
}>\alpha \geq 1}^{L}$ product to obtain%
\begin{eqnarray}
\mathbb{E}\left[ x^{2}\left( L\right) \right] _{G} &=&\frac{1}{Z}\int \left(
\prod_{\alpha =1}^{L}d^{n}y_{\alpha }e^{-\frac{1}{2}\sum_{\alpha }\left\vert 
\overrightarrow{y}_{\alpha }\right\vert ^{2}}\right) \left\vert \left\langle
y|M_{L}^{(H)}\right\rangle \right\vert ^{2}  \notag \\
&&\times \left( 1+\sum_{p=1}^{M}\left( -1\right) ^{p}\sum_{\substack{ 
\mathnormal{distinct}\text{ sets} \\ \text{of }p\text{ }\left( \alpha
,\alpha \prime \right) \text{ pairs}}}e^{-\beta ^{2}\sum_{p\text{ }\left(
\alpha ,\alpha \prime \right) \text{ pairs}}y\cdot A_{\alpha ,\alpha
^{\prime }}y}\right)   \label{GF2}
\end{eqnarray}%
%
%
\begin{equation}
Z=\int \left( \prod_{\alpha =1}^{L}d^{n}y_{\alpha }e^{-\frac{1}{2}%
\sum_{\alpha }\left\vert \overrightarrow{y}_{\alpha }\right\vert
^{2}}\right) \left( 1+\sum_{p=1}^{M}\left( -1\right) ^{p}\sum_{\substack{ 
\mathnormal{distinct}\text{ sets} \\ \text{of }p\text{ }\left( \alpha
,\alpha \prime \right) \text{ pairs}}}e^{-\beta ^{2}n\sum_{p\text{ }\left(
\alpha ,\alpha \prime \right) \text{ pairs}}y\cdot A_{\alpha ,\alpha
^{\prime }}y}\right)   \label{GF3}
\end{equation}%
where $M=\frac{L\left( L-1\right) }{2}$ and the matrices $A_{\alpha ,\alpha
^{\prime }}$ are a set of $M$ matrices that connect the space variables $y$
corresponding to the steps $\alpha $ and $\alpha ^{\prime }$. As $nL\times nL
$ matrices their elements are%
\begin{equation}
\left( A_{\alpha ,\alpha ^{\prime }}\right) _{i\beta ,j\beta ^{\prime
}}=\delta _{ij}\left( M_{\alpha }^{(H)}\left( \beta \right) -M_{\alpha
^{\prime }}^{(H)}\left( \beta \right) \right) \left( M_{\alpha }^{(H)}\left(
\beta ^{\prime }\right) -M_{\alpha ^{\prime }}^{(H)}\left( \beta ^{\prime
}\right) \right)   \label{GF4}
\end{equation}

In (\ref{GF2}) and (\ref{GF3}) the integrals are Gaussian, leading to%
\begin{eqnarray}
\mathbb{E}\left[ x^{2}\left( L\right) \right] _{G} &=&\frac{1}{Z}\frac{d}{%
d\gamma }\left( \det \left( \boldsymbol{1}-2B\gamma \right) ^{-\frac{1}{2}%
}\right.  \notag \\
&&\left. \left. +\sum_{p=1}^{M}\left( -1\right) ^{p}\sum_{\substack{ 
\mathnormal{distinct}\text{ sets}  \\ \text{of }p\text{ }\left( \alpha
,\alpha \prime \right) \text{ pairs}}}\det \left( \boldsymbol{1-}2B\gamma
+2\beta ^{2}n\sum_{p\text{ }\left( \alpha ,\alpha \prime \right) \text{ pairs%
}}A_{\alpha ,\alpha ^{\prime }}\right) ^{-\frac{1}{2}}\right) \right\vert
_{\gamma =0}  \label{GF5}
\end{eqnarray}%
and%
\begin{equation}
Z=1+\sum_{p=1}^{M}\left( -1\right) ^{p}\sum_{\substack{ \mathnormal{distinct}%
\text{ sets}  \\ \text{of }p\text{ }\left( \alpha ,\alpha \prime \right) 
\text{ pairs}}}\det \left( \boldsymbol{1}+2\beta ^{2}\sum_{p\text{ }\left(
\alpha ,\alpha \prime \right) \text{ pairs}}A_{\alpha ,\alpha ^{\prime
}}\right) ^{-\frac{1}{2}}  \label{GF6}
\end{equation}%
$B$ is the matrix%
\begin{equation*}
\left( B\right) _{i\beta ,j\beta ^{\prime }}=\delta _{ij}M_{L}^{(H)}\left(
\beta \right) M_{L}^{(H)}\left( \beta ^{\prime }\right)
\end{equation*}%
These closed form expressions are combinatorially complex, but may be used
to check the results obtained directly from the integrals in (\ref{GF0}) (%
\ref{GF0a}), at least for small $L$ values.

\subsection{The step factor}

Results for the step Gibbs factor are also relevant in particular because,
as discussed before, this factor is essentially identical to the Edwards
factor. It is also what is used in most Monte Carlo calculations 
\begin{equation}
e^{-\phi _{step}\left( z\right) }=\theta \left( \left\vert z\right\vert
^{2}-\varepsilon ^{2}\right)   \label{SF1}
\end{equation}%
$\theta $ being the Heaviside function. Using (\ref{2.7}) and (\ref{2.8})
one writes%
\begin{equation}
\mathbb{E}\left[ x^{2}\left( L\right) \right] _{S}=\frac{1}{Z}\int \left(
\prod_{\alpha =1}^{L}d^{n}y_{\alpha }e^{-\frac{1}{2}\sum_{\alpha }\left\vert 
\overrightarrow{y}_{\alpha }\right\vert ^{2}}\right) \left\vert \left\langle 
\overrightarrow{y}|M_{L}^{(H)}\right\rangle \right\vert ^{2}\prod_{\alpha
^{\prime }>\alpha \geq 1}^{L}\theta \left( \left\vert \left\langle
y|M_{\alpha }^{(H)}\right\rangle -\left\langle y|M_{\alpha ^{\prime
}}^{(H)}\right\rangle \right\vert ^{2}-\varepsilon ^{2}\right)   \label{SF2}
\end{equation}%
\begin{equation*}
\mathbb{Z}=\int \left( \prod_{\alpha =1}^{L}d^{n}y_{\alpha }e^{-\frac{1}{2}%
\sum_{\alpha }\left\vert \overrightarrow{y}_{\alpha }\right\vert
^{2}}\right) \prod_{\alpha ^{\prime }>\alpha \geq 1}^{L}\theta \left(
\left\vert \left\langle y|M_{\alpha }^{(H)}\right\rangle -\left\langle
y|M_{\alpha ^{\prime }}^{(H)}\right\rangle \right\vert ^{2}-\varepsilon
^{2}\right) 
\end{equation*}%
$\theta $ being the Heaviside function.

Here one obtains truncated Gaussian integrals: Given the correspondence $%
\beta \sim \frac{1}{\varepsilon }$, discussed before, we expect results
similar to those of the Gaussian kernel, in particular also a dependency on
the size of the cut-off $\varepsilon $. For example for $H=0.5$ we obtain
for the $\nu $ exponent

\begin{center}
$%
\begin{array}{lrr}
& n=2 & n=3 \\ 
\varepsilon =0.1 & 0.54 &  \\ 
\varepsilon =0.33 & 0.64 &  \\ 
\varepsilon =0.4 & 0.69 & 0.52 \\ 
\varepsilon =0.7 &  & 0.55 \\ 
\varepsilon =0.85 &  & 0.58%
\end{array}%
$
\end{center}

Monte Carlo simulations with Brownian or fractional processes and a
condition of minimal separation $\varepsilon $ for the polymer evolution, is
equivalent to the step factor (\ref{SF1}). Therefore the calculation of the
integrals in (\ref{SF2}) may be used as an alternative to Monte Carlo
simulations.

\textbf{Acknowledgments}

Partially supported by Funda\c{c}\~{a}o para a Ci\^{e}ncia e a Tecnologia
(FCT), project UIDB/04561/2020: https://doi.org/10.54499/UIDB/04561/2020

\end{document}